\documentclass[aps,prl,reprint,superscriptaddress,longbibliography]{revtex4-2}

\usepackage{amsmath,amssymb,mathtools,bm}
\usepackage{xcolor}
\usepackage{booktabs}
\usepackage{microtype}
\usepackage{graphicx}
\usepackage[hidelinks]{hyperref}
\graphicspath{{./}}
\newcommand{\ii}{\mathrm{i}}
\newcommand{\ee}{\mathrm{e}}
\newcommand{\sech}{\operatorname{sech}}
\newcommand{\sgn}{\operatorname{sgn}}
\newcommand{\tr}{\operatorname{tr}}
\newcommand{\ND}{\mathrm{ND}}
\newcommand{\DD}{\mathrm{D}}
\newcommand{\FF}{\mathrm{F}}
\newcommand{\bs}{\mathrm{BS}}

\begin{document}
\title{Explicit Boundary-Entropy Deficits in the XX Spin Chain with arbitrary boundary fields}

\author{Yi Qiao}
\affiliation{Institute of Modern Physics, Northwest University,
	Xi'an 710127, China}
\affiliation{Peng Huanwu Center for Fundamental Theory,
	Xi'an 710127, China}

\author{Pei Sun}
\email{sunpei@nwu.edu.cn}
\affiliation{Institute of Modern Physics, Northwest University,
	Xi'an 710127, China}
\affiliation{Peng Huanwu Center for Fundamental Theory,
	Xi'an 710127, China}

\author{Junpeng Cao}
\email{junpengcao@iphy.ac.cn}
\affiliation{Beijing National Laboratory for Condensed Matter Physics,
	Institute of Physics, Chinese Academy of Sciences,
	Beijing 100190, China}
\affiliation{School of Physical Sciences,
	University of Chinese Academy of Sciences,
	Beijing 100049, China}
\affiliation{Peng Huanwu Center for Fundamental Theory,
	Xi'an 710127, China}

\author{Wen-Li Yang}
\email{wlyang@nwu.edu.cn}
\affiliation{Institute of Modern Physics, Northwest University,
	Xi'an 710127, China}
\affiliation{Fundamental Discipline Research Center for Quantum Science
	and Technology of Shaanxi Province, Xian 710127, China}
\affiliation{Peng Huanwu Center for Fundamental Theory,
	Xi'an 710127, China}
\affiliation{Shaanxi Key Laboratory for Theoretical Physics Frontiers,
	Xi'an 710127, China}

\author{Yupeng Wang}
\email{yupeng@iphy.ac.cn}
\affiliation{Beijing National Laboratory for Condensed Matter Physics,
	Institute of Physics, Chinese Academy of Sciences,
	Beijing 100190, China}

\date{\today}

\begin{abstract}
The exact thermodynamics of the XX spin-1/2 chain with arbitrary Hermitian boundary fields is constructed via fusion techniques.
The rigorous result shows that (1) each end with a nonzero transverse field contributes $+\frac12T\ln2$ to the low-temperature free energy, whereas strictly diagonal and free ends contribute no such linear term; (2) each nondiagonal end has a characteristic temperature $T_B$, which scales as $T_B\propto h_\perp^2/J$ in the weak-field regime.
For $T\ll\min\{J,T_B\}$, the boundary entropy approaches a deficit of $\frac12\ln2$ relative to a diagonal end.
As the transverse field tends to zero, the boundary crossover scale vanishes and the low-temperature expansion becomes nonuniform.
The present result provides a clear benchmark not only for the boundary thermodynamics of open boundary quantum integrable systems, but also for dot defects in quantum critical systems.

\end{abstract}
\maketitle

Like Kondo impurities, local boundary fields can change the low-energy behavior of a critical system without changing its bulk Hamiltonian \cite{Affleck1998}.  At a boundary fixed point, the entropy $s_{\rm b}=\ln g$ fixes the linear-temperature term in the boundary free energy, with $g$ not necessarily an integer degeneracy \cite{AffleckLudwig1991}.
Such non-integer degeneracy also appears in spin-chain entanglement \cite{XavierRajabpour2020}.
Developing a microscopic finite-temperature description that connects these fixed-point results to lattice boundary thermodynamics beyond the low-energy conformal regime remains challenging. 
Obviously, the boundary $g$-function approaches unity as $T\to\infty$, since finite boundary fields do not alter the high-temperature state count.

The spectra of open XX \cite{Bilstein1999,Bilstein2000,Nepomechie2001} and XXZ chains \cite{Cao2003,Cao2013,Book2015}
have been studied extensively. Ground-state boundary energies have been calculated using root-of-unity solutions \cite{Murgan2007} and off-diagonal Bethe ansatz methods \cite{Liyuanyuan,Sun2019}. An alternative approach based on the zeros of transfer-matrix eigenvalues provides access to ground-state and surface energies \cite{Qiao2021,QiaoReview2023}.
Quantum-transfer-matrix methods have also yielded finite-temperature expressions for the boundary free energy with arbitrary boundary fields \cite{PozsgayRakos2018}, but the low-temperature asymptotics of the associated Fredholm determinants remain nontrivial.

In this paper, we take a simple but rather non-trivial model, i.e., the XX spin-$\frac12$ chain with arbitrary boundary fields as an example to study the thermodynamics of non-trivial boundaries. By solving the roots of the transfer matrix, we express the boundary free energy at all temperatures and for generic Hermitian boundary fields as a compact convergent series.
Its low-temperature expansion yields a simple counting law: the coefficient of $T$ is $\frac 12n_{\ND}\ln2$, where $n_{\ND}$ is the number of ends with a nonzero transverse field.  A free end is equivalent to a diagonal end for this counting.  The ground-state boundary energy and higher-order thermal corrections retain dependence on the boundary parameters, whereas at fixed nonzero transverse fields the linear coefficient depends only on the boundary class.  Thus each nondiagonal end carries an entropy deficit of $\frac12\ln2$ relative to a diagonal end.  The root representation also provides a microscopic account of this thermodynamic result: global physical-state selection, root densities, and endpoint contributions jointly determine the entropy constant and reorganize under strict boundary degeneration.

How can this discrete entropy classification be reconciled with a continuous change of the boundary field?  Our exact free energy shows that the boundary crossover scale vanishes with the transverse field, making the low-temperature expansion nonuniform.  At every fixed nonzero temperature, the free energy therefore approaches the strictly diagonal result continuously despite the discrete infrared entropy difference.  For a finite chain, the coefficient can be resolved only in a low-energy temperature window above the bulk level spacing and below the relevant boundary crossover scale.  The exact crossover function and finite-size calculations determine the visibility of this plateau.

\emph{Model and main result.---}
Our Hamiltonian is
\begin{equation}
H=J\sum_{j=1}^{N-1}(\sigma_j^x\sigma_{j+1}^x+\sigma_j^y\sigma_{j+1}^y)
+\boldsymbol h_-\cdot\boldsymbol\sigma_1
+\boldsymbol h_+\cdot\boldsymbol\sigma_N,
\label{eq:H}
\end{equation}
where $\sigma_j^\alpha (\alpha=x,y,z)$ indicate the Pauli matrices defined on site $j$. We set $J=k_B=1$ unless restoring an energy scale explicitly.  With $s_-=1$, $s_+=-1$, its integrable Hermitian boundary parametrization is
\begin{align}
h_\nu^z&=s_\nu\cot a_\nu\tanh b_\nu,\notag\\
(h_\nu^x,h_\nu^y)&=h_{\perp,\nu}(\cos\varphi_\nu,-\sin\varphi_\nu),\notag\\
h_{\perp,\nu}&\equiv\sqrt{(h_\nu^x)^2+(h_\nu^y)^2}
=\csc a_\nu\sech b_\nu.
\label{eq:Hb}
\end{align}
Here $\nu=-,+$ labels the left and right boundaries, respectively. We take $0<a_\nu<\pi$ and real $b_\nu,\varphi_\nu$, away from boundary-string thresholds and accidental degeneracies.  An end is nondiagonal ($\ND$) when $h_\perp\ne0$, diagonal ($\DD$) when $h_\perp=0$, and free ($\FF$) when its entire field vanishes.  The limit $b\to+\infty$ at fixed $a$ tends to $\DD$; then $a=\pi/2$ gives $\FF$.  The negative-$b$ limit is treated with the corresponding longitudinal-field sign.  Conventions and transfer-matrix normalization are given in Supplemental Material (SM) \cite{Supplemental}, Secs.~\ref{sm:conventions} and \ref{sm:construction}.

Let $X,Y\in\{\mathrm{ND},\mathrm{D},\mathrm{F}\}$ denote
the left and right boundary classes, respectively.
For $F_N^{X,Y}=-T\ln Z_N^{X,Y}$, define
\begin{equation}
F_{\rm b}^{X,Y}(T)=\lim_{N\to\infty}
\left[F_N^{X,Y}(T)-Nf_{\rm bulk}(T)\right],
\label{eq:Fbdef}
\end{equation}
and let $E_{\rm b}^{X,Y}$ be the analogous ground-state boundary energy.  The exact thermodynamic result is
\begin{equation}
F_{\rm b}^{X,Y}(T)=T\ln\cosh(2\beta)
-T[\mathcal B_X(\beta)+\mathcal B_Y(\beta)],
\label{eq:Fbexact}
\end{equation}
where $\beta=1/T$ and $\mathcal B_X=\sum_{p\in P_X}\widehat\Phi(p;\beta)$.  Each end carries its own parameters in the multisets $P_{\ND}=\{b,p(a)\}$, $P_{\DD}=\{p(a)\}$, and $P_{\FF}=\{0\}$, with $p(a)=\ii(\pi/2-a)$.  The normalized thermal function is
\begin{align}
\widehat\Phi(p;\beta)&=\sum_{m=1}^{\infty}
\ln\!\left[
\frac{\sqrt{1+q_m^2}+\cosh(2p)}{1+\cosh(2p)}
\right],\notag\\
q_m&=\frac{4\beta}{\pi(2m-1)}.
\label{eq:Phiexact}
\end{align}
This convergent series fixes $\widehat\Phi(p;0)=0$ and includes angular boundary strings by analytic continuation.  Equations~\eqref{eq:Fbexact} and \eqref{eq:Phiexact} hold for every $T>0$ in the stated parameter domain; their derivation is given below and in SM Sec.~\ref{sm:transforms}, Eqs.~\eqref{app:logZmaster} and \eqref{app:Phihat}--\eqref{app:thermalidentity}.

At fixed boundary parameters, their low-temperature expansion is
\begin{align}
F_{\rm b}^{X,Y}(T)={}&E_{\rm b}^{X,Y}
+\frac{n_{\ND}(X,Y)}2T\ln2\notag\\
&-\frac{\pi}{48}C_{X,Y}T^2+o(T^2),
\label{eq:main-law}
\end{align}
where $n_{\ND}=\bm 1_{X=\ND}+\bm 1_{Y=\ND}$ and $C_{X,Y}=\sum_{p\in P_X\sqcup P_Y}\cosh(2p)$.  Thus the nine ordered combinations reduce to three classes:
\begin{center}
\begin{tabular}{@{}lcc@{}}
\toprule
Ends & $n_{\ND}$ & $\displaystyle\lim_{T\to0}(F_{\rm b}-E_{\rm b})/T$\\
\midrule
$\ND$--$\ND$ & 2 & $\ln2$\\
$\ND$--$(\DD\text{ or }\FF)$ & 1 & $\frac12\ln2$\\
$(\DD\text{ or }\FF)$--$(\DD\text{ or }\FF)$ & 0 & $0$\\
\bottomrule
\end{tabular}
\end{center}
The middle row includes left--right exchange.  Each positive $\frac12T\ln2$ term is an entropy \emph{deficit}, $\Delta s_{\rm b}=-\frac12\ln2$, relative to a diagonal end.  At fixed boundary parameters, the linear coefficient depends only on the boundary class, whereas the quadratic coefficient depends on $a_\nu$ and $b_\nu$.  The expansion is derived in SM Sec.~\ref{sm:expansions}, Eq.~\eqref{app:lowTb}.

In particular, comparing $\ND(b,a)$ with its strict limiting reference $\DD(a)$ while keeping the other end $Y$ fixed gives
\begin{align}
F_{\rm b}^{\ND,Y}-F_{\rm b}^{\DD,Y}
&=-T\widehat\Phi(b;1/T)\notag\\
&=\mathfrak e(b)+\tfrac12T\ln2
\notag\\
&\quad-\frac{\pi}{48}\cosh(2b)T^2+o(T^2),
\label{eq:relative-law}
\end{align}
where $\mathfrak e(b)=-4b/[\pi\sinh(2b)]$, understood by continuity at $b=0$.  The expansion holds as $T\to0$ at fixed $b$; its growing quadratic coefficient signals a narrowing low-temperature regime for weak transverse fields.

\emph{Microscopic derivation.---}
We obtain Eq.~\eqref{eq:H} from the double-row transfer matrix of the open XXZ chain with generic reflection matrices in the Hermitian-field regime \cite{Sklyanin1988,deVega1994}.  Throughout, we specialize to the free-fermion point $\eta=\ii\pi/2$ ($\Delta=0$), where the fusion hierarchy truncates to a quadratic functional relation \cite{Nepomechie2001,Nepomechie2003}.  Removing the common state-independent factor $\sinh(2u)$ gives a regular reduced transfer matrix whose logarithmic derivative generates $H$ (SM Sec.~\ref{sm:construction}).  Evaluating the functional relation at an ordinary zero in the homogeneous chain yields
\begin{equation}
\begin{aligned}
&A_2(z)X(z)^2+A_1(z)X(z)+A_0(z)=0,\\
&X(z)=(-1)^N\coth^{2N}z.
\end{aligned}
\label{eq:zeroquadratic}
\end{equation}
The coefficients and functional-relation proof are in SM Sec.~\ref{sm:zeroequation}, Eqs.~\eqref{app:FR} and \eqref{app:Acoefs}.  Its solutions form allowed zero quartets under periodicity and crossing
\begin{equation}
\mathcal Q(z)=\{z,z+\ii\pi/2,-z-\ii\pi/2,-z\}.
\label{eq:quartet}
\end{equation}
One eigenvalue chooses one of the two crossing pairs from each quartet.  For the line modes, the choices give
\begin{equation}
E(\{s_j\})=\sum_{j=1}^{M}s_j\varepsilon_j,
\qquad \varepsilon(x)=2\sech(2x),\quad s_j=\pm1.
\label{eq:modeenergy}
\end{equation}
For $x>0$, the pair on $\operatorname{Im}z=+\pi/4$ contributes $-\varepsilon(x)$ and thus has $s=-1$; the pair on $\operatorname{Im}z=-\pi/4$ has $s=+1$ (SM Sec.~\ref{sm:projection}, Eq.~\eqref{app:pairenergy}).  Boundary-string modes are included as described in SM Sec.~\ref{sm:densities}.

\begin{figure*}[t]
	\centering
	\includegraphics[width=0.93\textwidth]{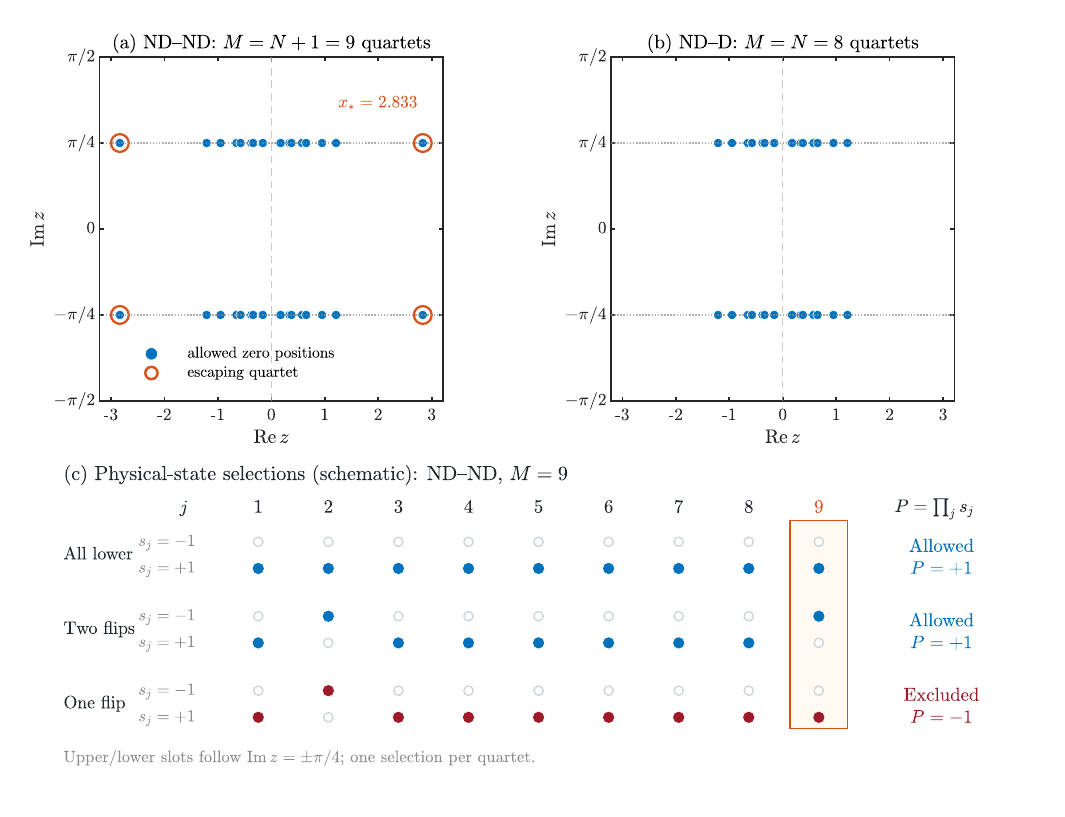}
	\caption{Allowed zeros and physical-state selection at $N=8$.  (a) $\ND$--$\ND$ at $b_+=4$, with $N+1=9$ allowed quartets.  Orange rings identify the additional quartet at $x_*=2.833$.  (b) The strict $\ND$--$\DD$ chain has $N=8$ quartets.  Each eigenvalue selects one crossing pair per quartet.  (c) Schematic pair selections, with horizontal spacings suppressed.  Upper and lower slots represent pairs on $\operatorname{Im}z=+\pi/4$ ($s_j=-1$) and $-\pi/4$ ($s_j=+1$), respectively.  For the displayed parameters, $P=\prod_js_j=\kappa=+1$, allowing an even number of upper slots: the two-flip configuration ($j=2,9$) is allowed, whereas the one-flip configuration is excluded.  For $\kappa=-1$, the allowed number is odd.  The orange column marks the additional quartet; the global restriction is absent in (b).  Parameters: $a_-=1.3$, $a_+=1.1$, $b_-=0.7$, $\varphi_-=0.1$, $\varphi_+=0.6$.}
	\label{fig:quartet}
\end{figure*}

Two nondiagonal ends supply $M=N+1$ modes with the physical constraint
\begin{equation}
\prod_j s_j=\kappa,\qquad
\kappa=\sgn\cos(\varphi_--\varphi_+).
\label{eq:kappa}
\end{equation}
Comparing the normalized zero-factor product with the transfer-matrix asymptotic \cite[Eq.~(5.3.14)]{Book2015} fixes the absolute sign in Eq.~\eqref{eq:kappa} (SM Sec.~\ref{sm:projection}, Eqs.~\eqref{app:parityasymptotic}--\eqref{app:absoluteparity}).  At orthogonal azimuths, the two projected spectra share a common zero-mode limit.  With one strictly diagonal end, the $N$ modes are independent.  Both constructions yield $2^N$ physical states, with partition functions
\begin{align}
Z_{\ND,\ND}&=\frac12\Bigg[\prod_{j=1}^{N+1}2\cosh(\beta\varepsilon_j)\notag\\
&\hspace{8mm}+\kappa(-1)^{N+1}\prod_{j=1}^{N+1}2\sinh(\beta\varepsilon_j)\Bigg],
\label{eq:ZNN}\\
Z_{\ND,\DD}&=\prod_{j=1}^{N}2\cosh(\beta\varepsilon_j).
\label{eq:ZND}
\end{align}
Figure~\ref{fig:quartet}(c) illustrates the global selection rule.  For $N=8$, the spectrum obtained from the finite-size zero equation agrees with independent spin-chain diagonalization to $6.04\times10^{-14}$; comparisons for all boundary classes are given in SM Sec.~\ref{sm:validation}, Fig.~\ref{fig:spectrumcheck}.  The thermodynamic boundary term additionally involves the $O(1)$ root density and endpoint contributions.

At fixed $T>0$, taking $N\to\infty$ makes the product of $\tanh(\beta\varepsilon_j)$ in Eq.~\eqref{eq:ZNN} vanish independently of $\kappa$; the projector leaves $-\ln2$ in $\ln Z$.  For a smooth test function with finite $f(\infty)$, the continuous-mode sum has the form
\begin{align}
\sum_{\rm cont}^{X,Y}f(x_j)={}&N\!\int_0^\infty\rho_0f
+\!\int_0^\infty\rho_1^{X,Y}f\notag\\
&-f(0)+c_{X,Y}f(\infty)+o(1),
\label{eq:sumrulemain}
\end{align}
where $\rho_0=4\sech(2x)/\pi$ and $c_{\ND,\ND}=0$, $c_{\ND,\DD}=-\frac12$, $c_{\DD,\DD}=0$.  The strict-class densities, discrete strings, and endpoint rules are derived in SM Sec.~\ref{sm:densities}, Eqs.~\eqref{app:rhoNN}--\eqref{app:lastdensity}.  For $f=\varepsilon$, $f(\infty)=0$; for $f=\ln[2\cosh(\beta\varepsilon)]$, it is $\ln2$.  An endpoint invisible to the ground-state energy can therefore affect entropy.

Combining these contributions gives the exact thermodynamic expression
\begin{equation}
\ln Z_{X,Y}=NA_0(\beta)+\mathcal B_X(\beta)+\mathcal B_Y(\beta)
-\ln\cosh(2\beta)+o(1).
\label{eq:mastermain}
\end{equation}
Here $A_0=\ln2+2\widehat\Phi(0;\beta)$ and the blocks are those defined above Eq.~\eqref{eq:Phiexact}.  Bulk subtraction gives Eq.~\eqref{eq:Fbexact}, including its normalization.  At fixed $p$,
\begin{align}
\widehat\Phi(p;\beta)={}&-\beta\mathfrak e(p)-\tfrac12\ln2
+\frac{\pi\cosh(2p)}{48\beta}\notag\\
&+o(\beta^{-1}),\qquad
\mathfrak e(p)=-\frac{4p}{\pi\sinh(2p)}.
\label{eq:Philow}
\end{align}
The $2+n_{\ND}$ blocks supply $-(2+n_{\ND})\ln2/2$, whereas $-\ln\cosh(2\beta)$ supplies $+\ln2$.  Their sum proves Eq.~\eqref{eq:main-law}; retaining the $\beta^{-1}$ terms gives its quadratic coefficient.  In Eq.~\eqref{eq:relative-law}, the opposite-end and angular blocks cancel exactly, leaving only $\widehat\Phi(b;\beta)$.
Equation~\eqref{eq:mastermain} combines the global projection and endpoint corrections into additive boundary blocks.

Strict diagonal degeneration changes the degree of the reduced eigenvalue:
\begin{equation}
\deg\widetilde\Lambda_{\ND,\ND}=2N+2
\quad\longrightarrow\quad
\deg\widetilde\Lambda_{\ND,\DD}=2N.
\label{eq:degrees}
\end{equation}
One complete allowed quartet escapes to infinity and the global selection constraint disappears, preserving the physical dimension.  Figure~\ref{fig:quartet} displays this finite-size reorganization.  At thermodynamic order, the moving density is
\begin{align}
\rho_{\rm esc}(x;b)&=\frac{\sech2(x+b)+\sech2(x-b)}\pi,\notag\\
\int_0^\infty\rho_{\rm esc}f&\xrightarrow{b\to\infty}\tfrac12 f(\infty).
\label{eq:rhoescmain}
\end{align}
The limiting smooth weight, together with removal of the complete quartet, accounts for the one-diagonal half-endpoint in Eq.~\eqref{eq:sumrulemain}.  When the remaining nondiagonal end is also removed, $\rho_1^{\ND,\DD}=\rho_1^{\DD,\DD}+\rho_{\rm esc}$: its $+\frac12f(\infty)$ cancels the explicit $-\frac12f(\infty)$.  No right half-endpoint survives in the strict two-diagonal chain (SM Sec.~\ref{sm:densities}, Eq.~\eqref{app:lastdensity}).  A free end follows by setting $a=\pi/2$.  These strict limits are taken in the full density or finite-temperature expression before the low-temperature expansion.

\begin{figure*}[t]
\centering
\includegraphics[width=0.94\textwidth]{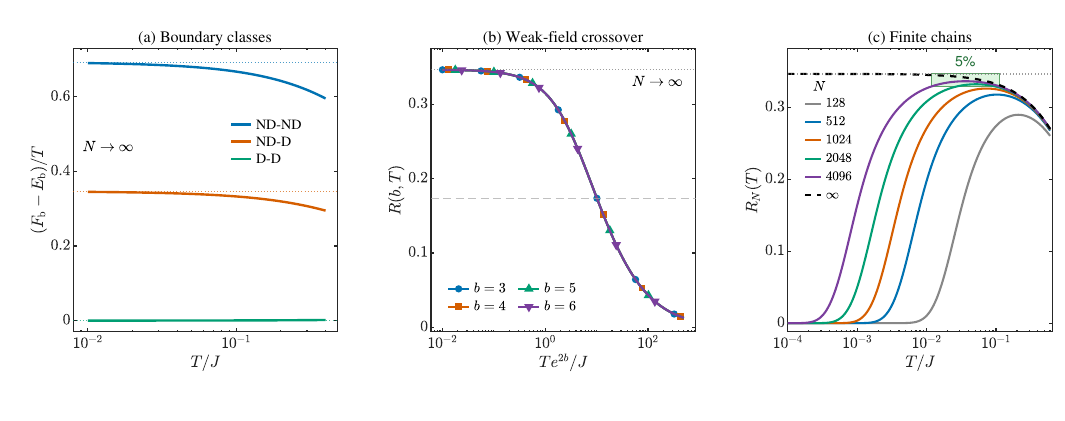}
\caption{Boundary free-energy coefficients and finite-size crossover.  (a) Exact thermodynamic coefficients for the three classes, with $a_-=1.3$, $a_+=0.2$, $b_-=0.7$, $b_+=0.6$ where applicable.  (b) Thermodynamic weak-field crossover $R(b,T)$ for $b=3$--$6$, with $T_{1/2}\ee^{2b}\simeq10.20$ in units $J=1$.  Panels (a,b) are evaluated in the thermodynamic limit; the low-energy regime is $T\ll J$.  The entropy difference corresponding to $R$ is $\Delta S_{\rm b}=-R-T\partial_TR$.  (c) Finite-chain $R_N(T)$ from ordinary-zero spectra for $\ND$--$\DD$ versus $\DD$--$\DD$, with $a_-=1.3$, $a_+=1.1$, $b_-=0.7$.  The black dashed curve gives the exact thermodynamic $R(b,T)$.  The shaded box marks the interval where the $N=4096$ curve lies within $5\%$ of $\frac12\ln2$.  All finite curves return to zero at sufficiently low $T$.}
\label{fig:numerics}
\end{figure*}

\emph{Crossover and finite-size visibility.---}
To resolve the coefficient of one end, compare $\ND(b,a)$ with its strict reference $\DD(a)$ and define
\begin{equation}
R(b,T)=\frac{F_{\rm b}^{\ND,Y}-F_{\rm b}^{\DD,Y}-\mathfrak e(b)}T
=-\widehat\Phi(b;1/T)-\frac{\mathfrak e(b)}T.
\label{eq:Rmain}
\end{equation}
Along this fixed-$a$ path, $h^z=s\cot a\tanh b$ approaches the reference value $s\cot a$.  The corresponding entropy difference is $\Delta S_{\rm b}=-R-T\partial_T R$, reducing to $-R$ on the infrared plateau.  Figure~\ref{fig:numerics} compares the thermodynamic coefficients and crossover with finite-chain results.

For weak transverse fields at fixed $a$, the thermodynamic weight in Eq.~\eqref{eq:rhoescmain} is centered at $x=|b|+O(1)$.  The corresponding mode energy sets the thermodynamic boundary scale
\begin{equation}
T_B\asymp2J\sech(2b)\asymp J\ee^{-2|b|}
\asymp h_\perp^2/J.
\label{eq:TB}
\end{equation}
Here $\asymp$ denotes equality up to finite factors, and the weak-field relations hold at fixed angle.  At fixed $b$, Eq.~\eqref{eq:relative-law} gives $R=\frac12\ln2-\pi T\cosh(2b)/48+o(T)$ in units $J=1$ (SM Eqs.~\eqref{app:Rkernel} and \eqref{app:Rlow}), so the infrared plateau requires $T\ll T_B$.  The uniform bound $R\le(4/\pi)\sqrt{E_*/T}$, with $E_*=2\sech(2b)$, holds for all real $b$ (SM Eq.~\eqref{app:Runiform}) and yields $R\to0$ as $T/E_*\to\infty$.  Thus $T_B\ll T\ll J$ is a diagonal-like low-energy regime for a weak but nonzero transverse field, whereas $T\gtrsim J$ probes the lattice high-temperature regime.

For a finite chain, resolving all nondiagonal ends additionally requires
\begin{gather}
\Delta_{\rm bulk}(N)\ll T\ll\min\{J,T_{B,\min}\},\notag\\
\Delta_{\rm bulk}(N)\asymp v_F/L=O(J/N).
\label{eq:window}
\end{gather}
Here $T_{B,\min}$ is the smallest scale among the nonzero transverse fields, in the absence of an additional anomalously small boundary scale.  The lower bound requires a thermal length $v_F/T\ll L$, with many bulk levels in the thermal window.  Boundary modes or near-degeneracies can lie below this bulk spacing.  At fixed $N$, a unique ground state gives $F_N=E_{0,N}+O(T\ee^{-\Delta_N/T})$, without a linear term (SM Sec.~\ref{sm:window}, Eqs.~\eqref{app:fixedN}--\eqref{app:finitewindow}).

The finite-chain counterpart is
\begin{equation}
R_N(T)=\frac{F_N^{\ND,Y}(T)-F_N^{\DD,Y}(T)
-[E_{0,N}^{\ND,Y}-E_{0,N}^{\DD,Y}]}T.
\label{eq:RNmain}
\end{equation}
The subtraction uses the ground-state energies of the same finite chains.  We solve the ordinary-zero equations for $\ND$--$\DD$ and $\DD$--$\DD$, at $a_-=1.3$, $a_+=1.1$, $b_-=0.7$, for nine sizes up to $N=4096$.  Figure~\ref{fig:numerics}(c) shows their approach to the thermodynamic curve and their return to zero at fixed-size low temperature.  At $N=4096$, $R_N$ reaches $0.336358$, within $2.95\%$ of $\frac12\ln2$, near $T=0.0363$.  It remains within $5\%$ over $0.0118\lesssim T\lesssim0.1119$; no such interval exists at $N=1024$.  This temperature-dependent interval quantifies the finite-size resolution of the infrared coefficient.  Independent Majorana spectra reproduce $R_N(T)$ with a maximum absolute difference below $3.51\times10^{-9}$ across all tested sizes and temperatures (SM Sec.~\ref{sm:validation}).

The exact finite-temperature function is continuous in the strict diagonal limit, but the energy-subtracted infrared coefficient has noncommuting limits:
\begin{equation}
\lim_{b\to\infty}\lim_{T\to0}R(b,T)=\tfrac12\ln2,
\qquad
\lim_{T\to0}\lim_{b\to\infty}R(b,T)=0.
\label{eq:limits}
\end{equation}
The shrinking crossover scale reconciles these statements.

\emph{Discussion.---}
The exact boundary free energy connects the universal classification of low-energy boundary entropy to the microscopic lattice spectrum.
The discrete low-temperature entropy difference is compatible with a continuous finite-temperature free energy: as the transverse field vanishes, the boundary crossover scale collapses, pushing the nondiagonal plateau to lower temperatures where it is limited by the bulk finite-size level spacing.  %
\iffalse 
\textcolor{red}{The fixed-$a$ comparison and a scan of the transverse field at fixed longitudinal field and azimuth yield the same infrared entropy deficit and weak-field scaling $T_B\propto h_\perp^2/J$, each relative to its corresponding diagonal boundary. Boundary energies, scale prefactors, and finite-temperature corrections can depend on the path (SM Sec.~\ref{sm:window}).}%
\else 
An experimentally natural control path holds the longitudinal field and azimuth fixed while varying the transverse field, with a diagonal boundary at the same longitudinal field as the reference.  This path and the fixed-$a$, variable-$b$ path approach the same low-energy boundary classes, retaining the entropy difference and the weak-field scaling $T_B\propto h_\perp^2/J$.  Their boundary energies, scale prefactors, and finite-temperature corrections can differ (SM Sec.~\ref{sm:window}).
The low-temperature slope of the boundary free-energy difference provides a thermodynamic diagnostic of boundary class when the thermodynamic limit is taken first. For finite chains, this coefficient can be resolved within the window of Eq.~\eqref{eq:window}; the strict $T\to0$ slope vanishes when both ground states are nondegenerate. Its visibility therefore depends jointly on temperature, chain length, and boundary-field strength, not merely on whether the transverse field is nonzero.

\emph{Acknowledgments.---}
The financial supports from
the National Natural Science Foundation of China (Grant Nos. 12675006, 12547107, 12434006 and 12247103)

\onecolumngrid
\clearpage
\setcounter{section}{0}
\setcounter{equation}{0}
\setcounter{figure}{0}
\setcounter{secnumdepth}{2}
\renewcommand{\thesection}{\Roman{section}}
\renewcommand{\theequation}{S\arabic{equation}}
\renewcommand{\thefigure}{S\arabic{figure}}
\renewcommand{\theHequation}{S\arabic{equation}}
\renewcommand{\theHfigure}{S\arabic{figure}}

\section*{Supplemental Material}

This Supplemental Material presents the transfer-matrix construction, ordinary-zero spectrum, and thermodynamic derivation of the boundary-entropy law.  Section~\ref{sm:expansions} derives Eq.~\eqref{eq:main-law}, and Sec.~\ref{sm:window} establishes the crossover and finite-size window.

\medskip
\noindent\textbf{Contents}\par\smallskip
\noindent
\begin{tabular*}{\textwidth}{@{}l@{\extracolsep{\fill}}r@{}}
\hyperref[sm:conventions]{\ref*{sm:conventions}. Conventions and boundary parametrization}
&\pageref{sm:conventions}\\
\hyperref[sm:construction]{\ref*{sm:construction}. Transfer-matrix construction and normalization}
&\pageref{sm:construction}\\
\hyperref[sm:zeroequation]{\ref*{sm:zeroequation}. Functional relation and ordinary-zero equation}
&\pageref{sm:zeroequation}\\
\hyperref[sm:projection]{\ref*{sm:projection}. Physical-state selection and finite-size partition functions}
&\pageref{sm:projection}\\
\hyperref[sm:densities]{\ref*{sm:densities}. Root densities, endpoints, and strict degeneration}
&\pageref{sm:densities}\\
\hyperref[sm:transforms]{\ref*{sm:transforms}. Exact thermodynamic transforms}
&\pageref{sm:transforms}\\
\hyperref[sm:expansions]{\ref*{sm:expansions}. Boundary energy and low-temperature expansion}
&\pageref{sm:expansions}\\
\hyperref[sm:window]{\ref*{sm:window}. Crossover, order of limits, and finite-size window}
&\pageref{sm:window}\\
\hyperref[sm:freecheck]{\ref*{sm:freecheck}. High-temperature and free-boundary limits}
&\pageref{sm:freecheck}\\
\hyperref[sm:validation]{\ref*{sm:validation}. Numerical spectra and thermodynamics}
&\pageref{sm:validation}\\
\end{tabular*}
\medskip

\section{Conventions and boundary parametrization}
\label{sm:conventions}

We use Pauli matrices and $J=k_B=1$ unless units are restored.  Equation~\eqref{eq:Hb} defines the physical boundary fields, with transverse magnitude $h_\perp=\sqrt{(h^x)^2+(h^y)^2}$.  The labels $\ND$, $\DD$, and $\FF$ distinguish nonzero transverse field, zero transverse field, and zero total field.  For $b\to+\infty$ at fixed $a$, $h^z\to s\cot a$; a subsequent choice $a=\pi/2$ gives a free end.  All comparisons in the thermal difference use this limiting diagonal reference.

For every fixed $T>0$, let $F_N^{X,Y}=-T\ln Z_N^{X,Y}$.  The thermodynamic boundary free energy is
\begin{equation}
F_{\rm b}^{X,Y}(T)=\lim_{N\to\infty}
\left[F_N^{X,Y}(T)-Nf_{\rm bulk}(T)\right],
\label{app:Fbdef}
\end{equation}
where the bulk and boundary terms use the same Hamiltonian normalization, with $Z_N(\beta=0)=2^N$.  The low-temperature expansion is taken after the thermodynamic limit; its finite-size realization is discussed in Sec.~\ref{sm:window}.

Unless stated otherwise, $0<a_\nu<\pi$ and $b_\nu,\varphi_\nu\in\mathbb R$.  For the generic two-nondiagonal zero divisor we also require $\cos(\varphi_--\varphi_+)\ne0$; its exceptional zero-mode limit is discussed in Sec.~\ref{sm:projection}.  The threshold values $a_\nu=\pi/4,3\pi/4$ and accidental finite-size degeneracies are obtained by limits from generic parameters.  Boundary strings outside the central strip are included by analytic continuation as explained below.

\section{Transfer-matrix construction and normalization}
\label{sm:construction}

The construction uses the reflection-equation framework \cite{Cherednik1984,Sklyanin1988} and the general six-vertex boundary matrices \cite{deVega1993,deVega1994}.  At the free-fermion point, the six-vertex matrix is
\begin{equation}
R(u)=\begin{pmatrix}
\ii\cosh u&0&0&0\\
0&\sinh u&\ii&0\\
0&\ii&\sinh u&0\\
0&0&0&\ii\cosh u
\end{pmatrix}.
\label{app:R}
\end{equation}
The left reflection matrix has entries
\begin{align}
K^-_{11}(u)&=2[\ii\sin a_-\cosh b_-\cosh u
+\cos a_-\sinh b_-\sinh u],\notag\\
K^-_{22}(u)&=2[\ii\sin a_-\cosh b_-\cosh u
-\cos a_-\sinh b_-\sinh u],\notag\\
K^-_{12}(u)&=\ee^{\ii\varphi_-}\sinh(2u),\qquad
K^-_{21}(u)=\ee^{-\ii\varphi_-}\sinh(2u),
\label{app:Km}
\end{align}
The right reflection matrix is
\begin{align}
K^+_{11}(u)&=2[\sin a_+\cosh b_+\sinh u
+\ii\cos a_+\sinh b_+\cosh u],\notag\\
K^+_{22}(u)&=2[\sin a_+\cosh b_+\sinh u
-\ii\cos a_+\sinh b_+\cosh u],\notag\\
K^+_{12}(u)&=\ee^{\ii\varphi_+}\sinh(2u),\qquad
K^+_{21}(u)=\ee^{-\ii\varphi_+}\sinh(2u).
\label{app:Kp}
\end{align}
With
\begin{equation}
T_0(u)=R_{0N}(u)\cdots R_{01}(u),\qquad
\widehat T_0(u)=R_{10}(u)\cdots R_{N0}(u),
\end{equation}
the open-chain transfer matrix is
\begin{equation}
t(u)=\tr_0[K^+_0(u)T_0(u)K^-_0(u)\widehat T_0(u)].
\label{app:t}
\end{equation}
The Yang--Baxter and reflection equations imply $[t(u),t(v)]=0$.

Regularity and crossing give state-independent zeros at $u=0,-\ii\pi/2$.  Since each matrix element is a finite Laurent polynomial in $\ee^{2u}$,
\begin{equation}
t(u)=\sinh(2u)\widetilde t(u),
\qquad
\Lambda(u)=\sinh(2u)\widetilde\Lambda(u).
\label{app:reduced}
\end{equation}
The regular free-fermion specialization of the conventional XXZ logarithmic derivative is
\begin{equation}
H=\ii\widetilde t'(0)\widetilde t(0)^{-1},
\qquad
E_a=\ii\left.\partial_u\ln\widetilde\Lambda_a(u)\right|_{u=0},
\label{app:energy}
\end{equation}
which yields Eqs.~\eqref{eq:H} and \eqref{eq:Hb}.

For two nondiagonal ends,
\begin{equation}
\widetilde t(0)=\lambda_0^{\ND,\ND}\,\mathrm{id},
\qquad
\lambda_0^{\ND,\ND}
=-4\ii(-1)^N\sinh(\ii a_-)\cosh b_-\sinh(\ii a_+)\cosh b_+.
\label{app:lambdaNN}
\end{equation}
After scaling one boundary before $b\to\infty$,
\begin{align}
\lambda_0^{\ND,\DD}
&=-2\ii(-1)^N\sinh(\ii a_-)\cosh b_-\sinh(\ii a_+),\notag\\
\lambda_0^{\DD,\ND}
&=-2\ii(-1)^N\sinh(\ii a_-)\sinh(\ii a_+)\cosh b_+.
\label{app:lambda1D}
\end{align}

\section{Functional relation and ordinary-zero equation}
\label{sm:zeroequation}

The free-fermion functional relation specializes the root-of-unity hierarchy \cite{Nepomechie2001,Nepomechie2002,Nepomechie2003,MurganNepomechieShi2006}.  Related spectral approaches include off-diagonal Bethe ansatz \cite{Cao2013,Book2015}, separation of variables \cite{Faldella2014,Kitanine2014}, modified algebraic Bethe ansatz \cite{Belliard2015,Avan2015}, and algebraic or coordinate constructions \cite{BaseilhacKoizumi2005,Crampe2011,Chernyak2023}.  Here we use ordinary transfer-matrix zeros.

Define the inhomogeneous bulk polynomial
\begin{equation}
A_\theta(u)=\prod_{l=1}^{N}
\cosh(u-\theta_l)\cosh(u+\theta_l).
\label{app:Atheta}
\end{equation}
For two nondiagonal boundaries,
\begin{align}
B(u)&=\prod_{\nu=\pm}\sinh(u-\ii a_\nu)\cosh(u-b_\nu),\notag\\
C(u)&=\prod_{\nu=\pm}\cosh(u+\ii a_\nu)\sinh(u+b_\nu),
\label{app:BC}
\end{align}
and
\begin{equation}
\chi=\frac18\left\{
\cosh\!\left[2(b_-+b_++\ii a_-+\ii a_+)\right]
+(-1)^N\cos\!\left[2(\varphi_--\varphi_+)\right]
\right\}.
\label{app:chi}
\end{equation}
Set
\begin{equation}
\Gamma(u)=-\ii\,
\frac{B(u)A_\theta(u)-C(u)A_\theta(u-\ii\pi/2)}{\cosh(2u)}.
\end{equation}
At the free-fermion point the apparent poles are removable, and the fusion hierarchy truncates to
\begin{equation}
\Lambda(u)\Lambda(u-\ii\pi/2)=16\sinh^2(2u)
\left[-\Gamma(u)\Gamma(u-\ii\pi/2)
+\chi\sinh^2(2u)A_\theta(u)A_\theta(u-\ii\pi/2)\right].
\label{app:FR}
\end{equation}

Dividing Eq.~\eqref{app:FR} by $16\sinh^2(2u)$ reduces the proof to interpolation.  The resulting functions are invariant under $u\to u+\ii\pi/2$ and $u\to-u-\ii\pi/2$ and belong to the $(N+2)$-dimensional space
\begin{equation}
\mathcal V_N=\left\{H_0+\sum_{m=1}^{N+1}H_m(\ee^{4mu}+\ee^{-4mu})\right\}.
\end{equation}
They agree at the $N$ quantum-determinant points, at $u=0$, and in the leading coefficient.  Distinct inhomogeneities make the $2N+1$ paired interpolation zeros distinct after the leading coefficient is removed, proving Eq.~\eqref{app:FR}; the homogeneous limit is taken afterward.

Using $B(u-\ii\pi/2)=C(-u)$ and $C(u-\ii\pi/2)=B(-u)$,
the homogeneous limit of Eq.~\eqref{app:FR} at an ordinary zero becomes
\begin{equation}
[B(z)X(z)-C(z)][C(-z)-B(-z)X(z)]
=\chi\mathcal K(z)X(z),
\label{app:factorzero}
\end{equation}
where
\begin{equation}
X(z)=(-1)^N\coth^{2N}z,
\qquad
\mathcal K(z)=\sinh^2(2z)\cosh^2(2z).
\end{equation}
The quadratic coefficients in Eq.~\eqref{eq:zeroquadratic} are
\begin{align}
A_2(z)&=B(z)B(-z),\notag\\
A_1(z)&=\chi\mathcal K(z)-B(z)C(-z)-C(z)B(-z),\notag\\
A_0(z)&=C(z)C(-z).
\label{app:Acoefs}
\end{align}
The branch product is
\begin{equation}
R_+(z)R_-(z)=\frac{C(z)C(-z)}{B(z)B(-z)}.
\label{app:Rprod}
\end{equation}
The formal solutions $z=\pm\ii\pi/4$ introduced by multiplication with $\cosh^2(2z)$ are excluded by the original functional relation.

For the strict $\ND$--$\DD$ limit,
\begin{align}
B_{\ND,\DD}(u)&=\frac12\ee^{-u}
\sinh(u-\ii a_-)\cosh(u-b_-)\sinh(u-\ii a_+),\notag\\
C_{\ND,\DD}(u)&=\frac12\ee^u
\cosh(u+\ii a_-)\sinh(u+b_-)
\cosh(u+\ii a_+),\notag\\
\chi_{\ND,\DD}&=\frac1{16}\ee^{2(\ii a_-+b_-+\ii a_+)}.
\label{app:BC1D}
\end{align}
The opposite orientation follows by exchanging the boundary labels.

\section{Physical-state selection and finite-size partition functions}
\label{sm:projection}

The two crossing pairs associated with Eq.~\eqref{eq:quartet} are represented by
\begin{align}
q_+(u;z)&=\frac{\ii}{2}[\sinh(2u)-\sinh(2z)],\notag\\
q_-(u;z)&=\frac{\ii}{2}[\sinh(2u)+\sinh(2z)].
\end{align}
At $u=0$,
\begin{equation}
q_-(0;z)=-q_+(0;z).
\label{app:qflip}
\end{equation}
Their logarithmic derivatives give
\begin{equation}
e_+(z)=-\frac{2\ii}{\sinh(2z)},\qquad e_-(z)=-e_+(z).
\label{app:pairenergy}
\end{equation}
For $z=x+\ii\pi/4$, $e_+=-2\sech(2x)$ and $e_-=+2\sech(2x)$.
Modulo the period $\ii\pi$, the roots of $q_+$ are $\{x+\ii\pi/4,-x+\ii\pi/4\}$, whereas those of $q_-$ are $\{x-\ii\pi/4,-x-\ii\pi/4\}$.  Thus $q_+$ corresponds to $s=-1$ and $q_-$ to $s=+1$ in Eq.~\eqref{eq:modeenergy}; the factor label $\tau=\pm$ and energy sign $s$ are opposite.  The corresponding pair selections are shown in Fig.~\ref{fig:quartet}(c).

A reduced eigenvalue of degree $2M$ factorizes as
\begin{equation}
\widetilde\Lambda_a(u)=c_a\prod_{j=1}^{M}q_{\tau_{a,j}}(u;z_j).
\end{equation}
For generic $\ND$--$\ND$ boundaries, $c_a$ is state independent, and Eq.~\eqref{app:qflip} together with Eq.~\eqref{app:lambdaNN} forces an even number of pair flips relative to an admissible eigenvalue.

\emph{Absolute parity.---}
For generic $\ND$--$\ND$ boundaries, the transfer-matrix asymptotic \cite[Eq.~(5.3.14)]{Book2015}, together with the normalization in Eq.~\eqref{app:lambdaNN}, gives
\begin{equation}
F_a(u)\equiv\frac{\widetilde\Lambda_a(u)}{\widetilde\Lambda_a(0)}
\sim\left(-\frac{\ii}{4}\right)^M
\frac{\cos\delta}{D}\,\ee^{2Mu},
\qquad u\to+\infty ,
\label{app:parityasymptotic}
\end{equation}
where $M=N+1$, $\delta=\varphi_--\varphi_+$, and
$D=\sin a_-\sin a_+\cosh b_-\cosh b_+>0$.
At the free-fermion point, $q_\tau(u)/q_\tau(0)=1-\frac{\ii}{2}e_\tau\sinh(2u)$.
Writing $e_{\tau_{a,j}}=s_{a,j}\varepsilon_j$ with $\varepsilon_j>0$, the zero-factor representation becomes
\begin{equation}
F_a(u)=\prod_{j=1}^{M}
\left[1-\frac{\ii}{2}s_{a,j}\varepsilon_j\sinh(2u)\right].
\label{app:normalizedmodeproduct}
\end{equation}
Comparison of the leading coefficients yields
\begin{equation}
\prod_{j=1}^{M}s_{a,j}\varepsilon_j=\frac{\cos\delta}{D},
\qquad
\prod_{j=1}^{M}s_{a,j}=\sgn\cos\delta=\kappa .
\label{app:absoluteparity}
\end{equation}
This constraint includes both line modes and finite-size boundary strings.
At $\cos\delta=0$, the continuous limit of Eq.~\eqref{app:absoluteparity} contains a zero mode, whose sign reversal connects the two parity sectors without changing their spectra.  Their partition functions therefore coincide.

For one diagonal end, the leading asymptotic coefficient is a nonscalar operator.  For the $\DD$--$\ND$ orientation, define $S^z=\frac12\sum_j\sigma_j^z$ and the one-flip operators
\begin{align}
\mathsf B&=\ii\sum_{k=1}^N
\ee^{\frac{\ii\pi}4\sum_{l>k}\sigma_l^z}\sigma_k^-
\ee^{-\frac{\ii\pi}4\sum_{l<k}\sigma_l^z},\notag\\
\mathsf C&=\ii\sum_{k=1}^N
\ee^{-\frac{\ii\pi}4\sum_{l>k}\sigma_l^z}\sigma_k^+
\ee^{\frac{\ii\pi}4\sum_{l<k}\sigma_l^z}.
\label{app:BCflip}
\end{align}
The hatted operators $\widehat{\mathsf B},\widehat{\mathsf C}$ are obtained by interchanging $l>k$ and $l<k$ in Eq.~\eqref{app:BCflip}.  Direct expansion of the two monodromies gives
\begin{equation}
t_{\DD,\ND}(u)=\ee^{(2N+2)u}\mathcal C_{\DD,\ND}
+O(\ee^{(2N+1)u}),
\qquad [\mathcal C_{\DD,\ND},t_{\DD,\ND}(v)]=0,
\label{app:Ccomm}
\end{equation}
where $\mathcal C_{\DD,\ND}=\mathcal C^{\rm diag}+\mathcal C^{\rm flip}$ and
\begin{align}
\mathcal C^{\rm diag}={}&\frac{\ii^{N+1}}{2^{2N+1}}
\left[-\ee^{\ii a_-}\sinh(\ii a_+-b_+)\ee^{\ii\pi S^z}
+\ee^{-\ii a_-}\sinh(\ii a_++b_+)\ee^{-\ii\pi S^z}\right],
\label{app:Cdiag}\\
\mathcal C^{\rm flip}={}&\frac{\ii^N\ee^{3\ii\pi/4}}{2^{2N+1}}
\Big[-\ee^{\ii\varphi_++\ii a_-}\mathsf C\ee^{\ii\pi S^z/2}
+\ee^{\ii\varphi_+-\ii a_-}\ee^{-\ii\pi S^z/2}\widehat{\mathsf C}\notag\\
&\hspace{12mm}-\ee^{-\ii\varphi_++\ii a_-}\ee^{\ii\pi S^z/2}\widehat{\mathsf B}
+\ee^{-\ii\varphi_+-\ii a_-}\mathsf B\ee^{-\ii\pi S^z/2}\Big].
\label{app:Cflip}
\end{align}
The spin-raising and spin-lowering terms prove $\mathcal C_{\DD,\ND}\not\propto\mathrm{id}$ for a generic nondiagonal end; commutativity in Eq.~\eqref{app:Ccomm} follows by taking the leading coefficient of $[t(u),t(v)]=0$.  On a common eigenstate its eigenvalue is proportional to the factorization constant $c_a$, whose sign compensates a single pair flip in Eq.~\eqref{app:qflip}.  There is no analogous scalar-coefficient parity restriction; completeness of the $N$ independent pair choices follows from the counting argument below.  The $\ND$--$\DD$ orientation follows by left--right exchange.

The same asymptotics fixes the change of degree.  Leading bulk $R$ entries preserve the auxiliary index, whereas an off-diagonal boundary entry flips it.  With two nondiagonal ends the two boundary flips close under the auxiliary trace, giving degree $2N+4$ before the common factor is removed.  With one diagonal end the naive leading path has an unpaired flip and vanishes.  The diagonal boundary lowers the boundary degree by one; a compensating off-diagonal bulk entry lowers it by another.  The resulting unreduced degree is $2N+2$, so removal of $\sinh(2u)$ gives Eq.~\eqref{eq:degrees}.  At generic parameters, this is the loss of one crossing-pair factor, or one allowed quartet mode.

For generic nondegenerate parameters, the $2^N$ physical common eigenstates map injectively to the sign configurations in the complete zero-factor decomposition.  There are exactly $2^N$ configurations satisfying Eq.~\eqref{app:absoluteparity}, or $2^N$ unrestricted configurations with one diagonal end, so the respective sets are exhausted; degeneracies follow by continuity.  The physical projection therefore gives
\begin{equation}
Z_{\ND,\ND}
=\sum_{\{s_j=\pm1\}}\frac{1+\kappa\prod_{j=1}^{M}s_j}{2}
\exp\!\left[-\beta\sum_{j=1}^{M}s_j\varepsilon_j\right],
\qquad M=N+1.
\label{app:projectedsum}
\end{equation}
Using $\sum_s\ee^{-\beta s\varepsilon}=2\cosh(\beta\varepsilon)$ and
$\sum_s s\ee^{-\beta s\varepsilon}=-2\sinh(\beta\varepsilon)$ gives Eq.~\eqref{eq:ZNN}, including its $\kappa(-1)^M$ sign.  The strict one-diagonal sum has no projector and gives Eq.~\eqref{eq:ZND}.  At $\beta=0$ the counts are $2^{N+1}/2=2^N$ and $2^N$, respectively.

\section{Root densities, endpoints, and strict degeneration}
\label{sm:densities}

Zero-density methods have been applied to XXZ surface energies \cite{Qiao2021}, other integrable spin chains \cite{Qiao2020,Li2022,Lu2023}, and XXX inversion relations \cite{QiaoReview2023}.  The main-line roots satisfy $z=x+\ii\pi/4$.  Roots that remain off this line approach boundary zeros or poles in the thermodynamic limit.  Only the angular parameters generate interior boundary strings, in the domains
\begin{equation}
0<a<\frac\pi4,
\qquad\text{or}\qquad
\frac{3\pi}{4}<a<\pi,
\end{equation}
with thermodynamic representative $z_{\bs}=\ii\min(a,\pi-a)$ and limiting energy
\begin{equation}
\varepsilon_{\bs}(a)=\frac2{|\sin2a|}.
\end{equation}
At finite $N$, the boundary-string positions are determined by Eq.~\eqref{eq:zeroquadratic} and differ from their thermodynamic limits.

For each quadratic branch define
\begin{equation}
Y_\nu(x)=\frac{X(x+\ii\pi/4)}{R_\nu(x+\ii\pi/4)},
\qquad
Z_\nu(x)=\frac1{2\pi}\operatorname{unwrap}\arg Y_\nu(x).
\end{equation}
The line-root condition is $|Y_\nu|=1$ and $Z_\nu=I\in\mathbb Z$.
The branch density is
\begin{equation}
\rho_{\nu,N}(x)=\frac2\pi\sech(2x)
-\frac{1}{2\pi\ii N}\partial_x\ln R_\nu(x+\ii\pi/4).
\end{equation}
Using Eq.~\eqref{app:Rprod}, the total density has the form
\begin{equation}
\rho_{\rm tot,N}(x)=\rho_0(x)+\frac1N\rho_1(x),
\qquad
\rho_0(x)=\frac4\pi\sech(2x).
\label{app:rho0}
\end{equation}
For two nondiagonal ends,
\begin{equation}
\rho_1^{\ND,\ND}(x)=\frac1\pi\sum_{\epsilon=\pm}
\left[
\sech2(x+\epsilon b_-)+\sech2(x+\epsilon b_+)
-\sech2(x+\epsilon\ii a_-)-\sech2(x+\epsilon\ii a_+)
\right].
\label{app:rhoNN}
\end{equation}
With one diagonal end and remaining real parameter $b$,
\begin{equation}
\rho_1^{\rm 1D}(x)=\frac1\pi\sum_{\epsilon=\pm}
\left[
\sech2(x+\epsilon b)
-\sech2(x+\epsilon\ii a_-)
-\sech2(x+\epsilon\ii a_+)
\right].
\label{app:rho1D}
\end{equation}
The discriminant, $\chi$, and the azimuthal angles disappear from the total density; they affect branch assignments and finite-size structure but not the displayed thermodynamic order.

Euler--Maclaurin summation on the two branches gives the strict-class rules
\begin{align}
\sum_{\rm cont}^{\ND,\ND}f(x_j)
&=N\int_0^\infty\rho_0f+\int_0^\infty\rho_1^{\ND,\ND}f-f(0)+o(1),
\label{eq:sumNN}\\
\sum_{\rm cont}^{\ND,\DD}f(x_j)
&=N\int_0^\infty\rho_0f+\int_0^\infty\rho_1^{\ND,\DD}f-f(0)-\frac12f(\infty)+o(1),
\label{eq:sumND}\\
\sum_{\rm cont}^{\DD,\DD}f(x_j)
&=N\int_0^\infty\rho_0f+\int_0^\infty\rho_1^{\DD,\DD}f-f(0)+o(1).
\label{eq:sumDD}
\end{align}
These formulas refer to continuous modes; discrete boundary strings are included separately before the thermal transform.  They are the strict-class versions of Eq.~\eqref{eq:sumrulemain}.    The left endpoint contributes $-f(0)$ because the formal $I=0$ solution is spurious.  In the two-nondiagonal problem the two right-end fractional parts are complementary and cancel.  In the strict one-diagonal problem the moving kernel
\begin{equation}
\rho_{\rm esc}(x;b)=\frac1\pi[\sech2(x+b)+\sech2(x-b)]
\label{app:rhoesc}
\end{equation}
has the distributional limit
\begin{equation}
\lim_{b\to\infty}\int_0^\infty\rho_{\rm esc}(x;b)f(x)\,dx
=\frac12f(\infty).
\label{app:esclimit}
\end{equation}
Subtracting the complete escaping root gives the additional $-\frac12f(\infty)$ in Eq.~\eqref{eq:sumND}.

For the remaining nondiagonal end, Eq.~\eqref{app:rho1D} and its strict diagonal limit give
\begin{align}
\rho_1^{\DD,\DD}(x)&=-\frac1\pi\sum_{\epsilon=\pm}
\left[\sech2(x+\epsilon\ii a_-)+\sech2(x+\epsilon\ii a_+)\right],\notag\\
\rho_1^{\ND,\DD}(x;b)&=\rho_1^{\DD,\DD}(x)+\rho_{\rm esc}(x;b).
\label{app:lastdensity}
\end{align}
Substituting Eq.~\eqref{app:lastdensity} into Eq.~\eqref{eq:sumND} before taking $b\to\infty$ produces $+\frac12f(\infty)$ from the moving kernel.  It cancels the explicit $-\frac12f(\infty)$, proving Eq.~\eqref{eq:sumDD} and removing the last $\frac12T\ln2$.  The free boundary is the special diagonal point $a=\pi/2$, for which $p(a)=0$; hence the $\DD$--$\FF$ and $\FF$--$\FF$ rules follow directly from the same density and thermal transform.

\section{Exact thermodynamic transforms}
\label{sm:transforms}

Define
\begin{align}
q_m&=\frac{4\beta}{\pi(2m-1)},\qquad
p(a)=\ii\left(\frac\pi2-a\right),\notag\\
\widehat\Phi(p;\beta)
&=\sum_{m=1}^{\infty}
\ln\!\left(
\frac{\sqrt{1+q_m^2}+\cosh2p}{1+\cosh2p}
\right).
\label{app:Phihat}
\end{align}
The normalized boundary blocks are
\begin{align}
\mathcal B_{\ND}(b,a;\beta)&=\widehat\Phi(b;\beta)+\widehat\Phi[p(a);\beta],\notag\\
\mathcal B_{\DD}(a;\beta)&=\widehat\Phi[p(a);\beta],\qquad
\mathcal B_{\FF}(\beta)=\widehat\Phi(0;\beta),
\label{app:Bblocks}
\end{align}
and, through $O(N)+O(1)$,
\begin{equation}
\ln Z_{X,Y}=N A_0(\beta)+\mathcal B_X(\beta)+\mathcal B_Y(\beta)
-\ln\cosh(2\beta)+o(1),
\label{app:logZmaster}
\end{equation}
where $A_0=\ln2+2\widehat\Phi(0;\beta)$.  This normalization gives the absolute boundary term in Eq.~\eqref{eq:main-law} and the relative law in Eq.~\eqref{eq:relative-law}.

Let
\begin{equation}
h_\beta(x)=\ln[2\cosh(2\beta\sech2x)].
\end{equation}
The bulk transform is
\begin{equation}
A_0(\beta)=\int_0^\infty\rho_0(x)h_\beta(x)\,dx
=\ln2+2\widehat\Phi(0;\beta).
\end{equation}
Every boundary density uses the kernel
\begin{equation}
K_p(x)=\sech2(x+p)+\sech2(x-p).
\end{equation}
For real $b$,
\begin{equation}
\int_0^\infty K_b(x)h_\beta(x)\,dx
=\pi\left[\frac12\ln2+\widehat\Phi(b;\beta)\right].
\label{app:thermalidentity}
\end{equation}
For $p=p(a)=\ii(\pi/2-a)$, the physical answer is the analytic continuation from the no-boundary-string strip $\pi/4<a<3\pi/4$.  In a string domain the fixed-contour integral differs from this continuation by
\begin{equation}
\pi h_{\bs}(a;\beta),
\qquad
h_{\bs}(a;\beta)=\ln\left[2\cosh\left(\frac{2\beta}{|\sin2a|}\right)\right].
\end{equation}
The analytic continuation therefore incorporates the discrete boundary-string contribution in Eqs.~\eqref{app:Bblocks} and \eqref{app:logZmaster}.

\section{Boundary energy and low-temperature expansion}
\label{sm:expansions}

Boundary critical behavior \cite{Alcaraz1987} and boundary scattering \cite{Ghoshal1994,Grisaru1995} provide complementary perspectives on boundary physics.  Here we extract the energy and thermal coefficients directly from the normalized zero-density blocks.

The unified one-boundary ground-state blocks are
\begin{align}
\mathcal E_{\ND}(b,a)&=\mathfrak e(b)+\mathfrak e[p(a)],\notag\\
\mathcal E_{\DD}(a)&=\mathfrak e[p(a)],\qquad
\mathcal E_{\FF}=\mathfrak e(0)=-\frac2\pi,
\end{align}
with
\begin{equation}
\mathfrak e(p)=-\frac{4p}{\pi\sinh2p}.
\end{equation}
Thus
\begin{equation}
E_{\rm gs}^{X,Y}(N)=-\frac{4N}{\pi}+2+\mathcal E_X+\mathcal E_Y+o(1).
\end{equation}
Hence $e_{\rm bulk}=-4/\pi$ and $E_{\rm b}^{X,Y}=2+\mathcal E_X+\mathcal E_Y$ in the convention of Eq.~\eqref{app:Fbdef}.

At low temperature,
\begin{align}
A_0(\beta)&=\frac{4\beta}{\pi}+\frac{\pi}{24\beta}+o(\beta^{-1}),\notag\\
\widehat\Phi(p;\beta)
&=-\beta\mathfrak e(p)-\frac12\ln2
+\frac{\pi\cosh2p}{48\beta}+o(\beta^{-1}),\notag\\
-\ln\cosh(2\beta)&=-2\beta+\ln2+O(\ee^{-4\beta}).
\label{app:lowTblocks}
\end{align}
For a boundary of type $X$, introduce the multiset
\begin{equation}
P_{\ND}=\{b,p(a)\},\qquad
P_{\DD}=\{p(a)\},\qquad
P_{\FF}=\{0\}.
\end{equation}
Since $|P_X|=1+\bm 1_{X=\ND}$, the two-boundary count is
\begin{equation}
K_{X,Y}=|P_X|+|P_Y|=2+n_{\ND}(X,Y).
\end{equation}
Taking the thermodynamic limit before the low-temperature expansion of Eq.~\eqref{app:logZmaster} gives, through $O(N)+O(1)$,
\begin{equation}
F_N^{X,Y}(T)=E_{\rm gs}^{X,Y}(N)
+\frac{n_{\ND}}2T\ln2
-\frac\pi{24}\left[N+\frac12C_{X,Y}\right]T^2+o(T^2),
\label{app:lowTall}
\end{equation}
where
\begin{equation}
C_{X,Y}=\sum_{p\in P_X\sqcup P_Y}\cosh2p.
\end{equation}
After subtracting $Nf_{\rm bulk}(T)$ as in Eq.~\eqref{app:Fbdef}, Eq.~\eqref{app:lowTall} becomes
\begin{equation}
F_{\rm b}^{X,Y}(T)=E_{\rm b}^{X,Y}
+\frac{n_{\ND}}2T\ln2-\frac{\pi}{48}C_{X,Y}T^2+o(T^2),
\label{app:lowTb}
\end{equation}
which proves Eq.~\eqref{eq:main-law} with the required order of limits.

\section{Crossover, order of limits, and finite-size window}
\label{sm:window}

\emph{Boundary crossover.---}
For $b\ge0$, compare $\ND(b,a)$ with its limiting diagonal boundary $\DD(a)$ at fixed opposite boundary.  Their longitudinal fields are $s\cot a\tanh b$ and $s\cot a$, respectively.  With the corresponding signed reference, the thermal difference is even in $b$.  Define
\begin{equation}
R(b,T)=\frac{F_{\rm b}^{\ND,Y}(T)-F_{\rm b}^{\DD,Y}(T)-\mathfrak e(b)}{T}
=-\widehat\Phi(b;1/T)-\frac{\mathfrak e(b)}{T}.
\label{app:Rdef}
\end{equation}

\emph{Control at fixed physical longitudinal field.---}
An experimental scan can instead hold $h_z$ and the azimuth fixed while varying $h_\perp$, using the reference $(0,h_z)$.  Restoring $J$, this path satisfies $\cot a(b)=h_z/[sJ\tanh b]$ together with $h_\perp=J\csc a(b)\sech b$.  Both auxiliary parameters change for nonzero $h_z$; for $h_z=0$, $a=\pi/2$ remains fixed along the weak-field branch.  For a finite fixed $h_z$, $a(b)$ tends to a fixed interior value as $h_\perp\to0$, so the same weak-field relation $T_B\propto h_\perp^2$ and the same infrared entropy deficit follow.  At nonzero $h_z$ the angle blocks no longer cancel exactly against the diagonal reference at finite field, which changes the boundary energy and finite-temperature corrections.  Equation~\eqref{app:Rdef} and its positive-kernel representation below refer specifically to the fixed-$a$ comparison.

\emph{Free-energy coefficient versus entropy.---}
At fixed boundary parameters, write $\Delta F_{\rm b}=F_{\rm b}^{\ND,Y}-F_{\rm b}^{\DD,Y}$ and $\Delta S_{\rm b}=S_{\rm b}^{\ND,Y}-S_{\rm b}^{\DD,Y}$.  The entropy difference follows from $\Delta F_{\rm b}=\mathfrak e(b)+TR(b,T)$:
\begin{equation}
\Delta S_{\rm b}(b,T)
=-\partial_T\Delta F_{\rm b}(b,T)
=-R(b,T)-T\partial_T R(b,T).
\label{app:entropyvsR}
\end{equation}
Thus $\Delta S_{\rm b}\to-\frac12\ln2$ as $T\to0$ at fixed parameters, where $R\to\frac12\ln2$ and $T\partial_T R\to0$.  At finite temperature, the derivative term distinguishes the entropy difference from the free-energy ratio.

Equations~\eqref{app:thermalidentity} and \eqref{app:lowTblocks}, with
$\int_0^\infty K_b\,dx=\pi/2$ and
$\int_0^\infty K_b\varepsilon\,dx=-\pi\mathfrak e(b)$, give the exact thermodynamic representation
\begin{equation}
R(b,T)=\frac1\pi\int_0^\infty K_b(x)
G\!\left(\frac{\varepsilon(x)}T\right)\,dx,\qquad
G(y)=y-\ln\cosh y,\qquad
\varepsilon(x)=2\sech(2x).
\label{app:Rkernel}
\end{equation}
For $y\ge0$, $G(0)=0$, $G(y)\to\ln2$ as $y\to\infty$, and $0\le G(y)\le\ln2$.  Thus every fixed finite $b$ gives $R\to\frac12\ln2$ as $T\to0$.  More precisely,
\begin{equation}
R(b,T)=\frac12\ln2-\frac{\pi}{48}\cosh(2b)\,T+o(T)
\qquad (T\to0\ \text{at fixed }b).
\label{app:Rlow}
\end{equation}
The correction grows as the transverse field decreases, making the low-temperature expansion nonuniform as $|b|\to\infty$.

For weak transverse fields at fixed $a$, $|b|\gg1$, the kernel weight is concentrated at $x=|b|+O(1)$.  The thermal argument in Eq.~\eqref{app:Rkernel} therefore varies on the energy scale $\varepsilon(|b|)$, yielding
\begin{equation}
T_B\asymp\varepsilon(|b|)=2\sech(2b)
\asymp\ee^{-2|b|}\asymp h_\perp^2,\qquad
h_\perp=\csc a\,\sech b.
\label{app:TBscale}
\end{equation}
The prefactor of $T_B$ depends on the crossover convention, such as the midpoint temperature in Sec.~\ref{sm:validation}.  Equation~\eqref{app:Rlow} gives the plateau condition $T\ll T_B$.  At fixed $T>0$, the diagonal limit moves the kernel weight to $\varepsilon=0$, where $G(0)=0$, and hence yields $R\to0$, consistently with Eq.~\eqref{eq:limits}.

\emph{Uniform suppression for $T\gg E_*(b)$.---}
Define the boundary energy scale
\begin{equation}
E_*(b)=2\sech(2b).
\label{app:Estar}
\end{equation}
Even extension of the integrand in Eq.~\eqref{app:Rkernel}, followed by a shift, gives
\begin{equation}
R(b,T)=\frac1\pi\int_{-\infty}^{\infty}\sech(2u)\,
G\!\left[\frac{E_*(b)}{T}\frac{\cosh(2b)}{\cosh 2(u+b)}\right]du.
\label{app:Rshifted}
\end{equation}
For $y\ge0$, $G(y)\le\min(y,\ln2)\le y^\alpha$ for every $0<\alpha<1$.
Since $|d\ln\cosh(2x)/dx|\le2$, the ratio of coshes in Eq.~\eqref{app:Rshifted} is at most $\ee^{2|u|}$.  With $\sech(2u)\le2\ee^{-2|u|}$ this yields, uniformly in real $b$,
\begin{equation}
0\le R(b,T)\le\frac{2}{\pi(1-\alpha)}
\left[\frac{E_*(b)}{T}\right]^\alpha,\qquad 0<\alpha<1.
\label{app:Runiform}
\end{equation}
Taking $\alpha=1/2$ gives $R\le(4/\pi)\sqrt{E_*/T}$ and hence $R\to0$ uniformly in real $b$ as $T/E_*\to\infty$.  In the weak-field regime, $T_B\asymp E_*$, and the diagonal-like low-energy window is $T_B\ll T\ll J$.  For $E_*=O(J)$ this scale-separated window is absent; a large ratio then probes high-temperature insensitivity.

\emph{Fixed size versus thermodynamic low temperature.---}
For a fixed finite chain, let $g_0$ be the ground-state degeneracy and $\Delta_N>0$ the gap from that ground-state manifold to the first higher level.  Its low-temperature partition function and free energy obey
\begin{align}
Z_N(T)&=\ee^{-E_{0,N}/T}\left[g_0+O(\ee^{-\Delta_N/T})\right],\notag\\
F_N(T)&=E_{0,N}-T\ln g_0+O(T\ee^{-\Delta_N/T}).
\label{app:fixedN}
\end{align}
When $g_0=1$ there is no linear term.  If an exact finite-size zero mode gives $g_0>1$, its $-T\ln g_0$ is a finite-size ground-state entropy, not the thermodynamic boundary term in Eq.~\eqref{app:lowTb}.

\emph{Finite-chain observable for the boundary plateau.---}
For chains of equal length $N$ and a common opposite boundary $Y$, the finite-size counterpart of Eq.~\eqref{app:Rdef} is defined using the same boundary-parameter matching.  With $E_{0,N}^{X,Y}$ the lowest physical energy after any required projection,
\begin{align}
\Delta E_{0,N}
&=E_{0,N}^{\ND,Y}-E_{0,N}^{\DD,Y},\notag\\
R_N(b,T)
&=\frac{F_N^{\ND,Y}(T)-F_N^{\DD,Y}(T)-\Delta E_{0,N}}{T}.
\label{app:RNdef}
\end{align}
Relative to thermodynamic ground-state subtraction, using $\Delta E_{0,N}$ removes the term $[\Delta E_{0,N}-\mathfrak e(b)]/T$.  For fixed $T>0$, Eq.~\eqref{app:RNdef} tends to Eq.~\eqref{app:Rdef} as $N\to\infty$ and $\Delta E_{0,N}\to\mathfrak e(b)$.

Let $g_{0,N}^{\ND,Y}$ and $g_{0,N}^{\DD,Y}$ be the two ground-state
degeneracies, and let $\Delta_N^{\ND,Y}$ and $\Delta_N^{\DD,Y}$ be the
positive gaps above their complete ground-state manifolds.  Applying
Eq.~\eqref{app:fixedN} to each chain gives
\begin{equation}
R_N(b,T)
=-\ln\frac{g_{0,N}^{\ND,Y}}{g_{0,N}^{\DD,Y}}
+O\!\left(\ee^{-\Delta_N^{\ND,Y}/T}
          +\ee^{-\Delta_N^{\DD,Y}/T}\right),
\qquad T\to0\quad(N\text{ fixed}).
\label{app:RNfixedN}
\end{equation}
For nondegenerate ground states, $R_N\to0$ at fixed $N$; with degeneracies, the limit is the logarithm in Eq.~\eqref{app:RNfixedN}.  These finite-size limits differ from the thermodynamic half-logarithm.  The finite-size entropy difference is $\Delta S_N=-R_N-T\partial_T R_N$ at fixed parameters.

The finite-size correction separates into ground-state and thermal contributions.  With
$\Delta F_N=F_N^{\ND,Y}-F_N^{\DD,Y}$ and
$\Delta F_{\rm b}=F_{\rm b}^{\ND,Y}-F_{\rm b}^{\DD,Y}$, the exact identity
\begin{equation}
R_N-R=
-\frac{\Delta E_{0,N}-\mathfrak e(b)}{T}
+\frac{\Delta F_N-\Delta F_{\rm b}}{T}
\label{app:RNcorrection}
\end{equation}
shows that an $O(N^{-1})$ ground-state correction contributes $O((NT)^{-1})$ to $R_N-R$.  Resolving the thermodynamic half-logarithm requires both this finite-size correction and the deviation of $R(T)$ from $\frac12\ln2$ to be small.

The critical bulk-mode spacing sets the lower temperature scale.  From the free-chain spectrum in Sec.~\ref{sm:freecheck}, restoring the exchange coefficient $J$ and setting the lattice spacing and $\hbar$ to one gives the Jordan--Wigner occupation energy $\omega(k)=4J\cos k$, with $k_j=j\pi/(N+1)$.  The two-level energy in Eq.~\eqref{eq:modeenergy} is half this occupation-energy splitting.  Around $k_F=\pi/2$,
\begin{equation}
v_F=|\partial_k\omega(k_F)|=4J,\qquad
\delta k=\frac{\pi}{N+1},\qquad
\delta\omega\simeq v_F\delta k=O(J/N).
\label{app:bulkspacing}
\end{equation}
Generic local boundary fields shift reflection phases and level offsets while preserving this bulk scaling.  The bulk finite-size scale is therefore $\Delta_{\rm bulk}(N)\asymp v_F/L$, with $L\simeq N$.  Boundary modes, exact zero modes, and near-degeneracies may lie below this spacing.

The number of bulk modes thermally resolved near a Fermi point
is of order $LT/v_F$. Replacing their sum by a thermodynamic
integral requires this number to be large, or equivalently,
a thermal length $\xi_T\asymp v_F/T\ll L$.
For each nondiagonal boundary $\nu=\pm$, $T_{B,\nu}$ denotes
the crossover scale introduced in Eq.~\eqref{app:TBscale} with the corresponding
boundary parameters. Together with the low-energy and
boundary-crossover conditions, this gives, for $n_{\mathrm{ND}}\geq 1$,
\begin{equation}
\Delta_{\mathrm{bulk}}(N)\ll T\ll
\min\bigl\{J,T_{B,\min}\bigr\},
\qquad
T_{B,\min}
=\min_{\nu:\,h_{\perp,\nu}\neq 0}T_{B,\nu}.
\label{app:finitewindow}
\end{equation}
provided no additional anomalously small boundary scale intervenes.  For $n_{\ND}=0$ there is no transverse-field crossover condition, and the usual bulk window is $\Delta_{\rm bulk}\ll T\ll J$.  In the normalization $J=1$, the lower bound is the $N^{-1}\ll T$ condition stated in the Letter.  In the weak-field regime a well-separated plateau requires $N T_{B,\min}\gg1$ up to velocity factors.

\emph{Projection and order of limits.---}
For $\ND$--$\ND$, Eq.~\eqref{eq:ZNN} can be factorized as
\begin{equation}
Z_{\ND,\ND}
=\frac12\prod_{j=1}^{M}2\cosh(\beta\varepsilon_j)
\left[1+\kappa(-1)^M\prod_{j=1}^{M}\tanh(\beta\varepsilon_j)\right],
\qquad M=N+1.
\label{app:projectedfactor}
\end{equation}
At fixed finite $N$ with positive mode energies, the product of hyperbolic tangents tends to one as $T\to0$.  When $\kappa(-1)^M=+1$, the bracket tends to two and cancels the prefactor; when $\kappa(-1)^M=-1$, the leading term cancels, so the physical ground state flips the softest mode.  Both cases implement finite-size ground-state selection.  At fixed $T>0$ followed by $N\to\infty$, instead, arbitrarily soft bulk modes make the product tend to zero.  The bracket tends to one, leaving $-\ln2$ in $\ln Z$ before the density and endpoint terms are combined, independently of $\kappa$.

\section{High-temperature and free-boundary limits}
\label{sm:freecheck}

\emph{High-temperature normalization.---}
At high temperature,
\begin{align}
A_0(\beta)&=\ln2+\beta^2-\frac12\beta^4+O(\beta^6),\notag\\
\widehat\Phi(p;\beta)
&=\frac{\beta^2}{1+\cosh2p}
-\frac{2+\cosh2p}{3(1+\cosh2p)^2}\beta^4+O(\beta^6).
\end{align}
Define
\begin{equation}
S_1(X,Y)=\sum_{p\in P_X\sqcup P_Y}\frac1{1+\cosh2p}.
\label{app:highTcoeff}
\end{equation}
Substituting these expansions into Eq.~\eqref{app:logZmaster} yields,
through the thermodynamic $O(N)+O(1)$ terms,
\begin{equation}
\ln Z_{X,Y}
=N\ln2+[N-2+S_1(X,Y)]\beta^2+O(\beta^4)+o(1).
\label{app:highTcheck}
\end{equation}
The remainder $o(1)$ is the thermodynamic finite-size correction, while $O(\beta^4)$ contains bulk and boundary contributions.  The expansion recovers $Z_N(0)=2^N$ and the corresponding bulk and boundary trace moments.

\emph{Double-free finite chain.---}
For $\FF$--$\FF$, the ordinary-zero equation is a perfect square and reduces to
\begin{equation}
\tanh^{2N+2}z=(-1)^{N+1}.
\end{equation}
The exact single-particle energies are
\begin{equation}
\varepsilon_j=2\cos\frac{j\pi}{N+1},\qquad j=1,\ldots,N,
\end{equation}
and
\begin{equation}
\ln Z_{\FF,\FF}^{(N)}(\beta)=
\sum_{j=1}^{N}\ln\left[2\cosh\left(2\beta\cos\frac{j\pi}{N+1}\right)\right].
\label{app:ZFF}
\end{equation}
For odd $N$, the middle term is the zero-energy center mode and contributes $\ln2$.  Euler--Maclaurin summation of Eq.~\eqref{app:ZFF} gives Eq.~\eqref{app:logZmaster} with two free boundary blocks and no linear term in Eqs.~\eqref{app:lowTall} and \eqref{app:lowTb}.

In the order of limits used throughout the Letter, the resulting
double-free boundary term is
\begin{equation}
F_{\rm b}^{\FF,\FF}(T)
=E_{\rm b}^{\FF,\FF}-\frac{\pi}{24}T^2+o(T^2),
\qquad E_{\rm b}^{\FF,\FF}=2-\frac4\pi.
\label{app:FFlowTcheck}
\end{equation}
The absence of a linear term follows from taking the thermodynamic limit first.  At fixed odd $N$, the exact center mode retains its ground-state degeneracy, as discussed in Sec.~\ref{sm:window}.

\section{Numerical spectra and thermodynamics}
\label{sm:validation}

\emph{Finite-size spectra.---}
The line roots and boundary strings are obtained from the finite-size ordinary-zero equation~\eqref{eq:zeroquadratic}.  Their energies and the selection rule of Sec.~\ref{sm:projection} determine the complete physical spectrum.  The free--free spectrum follows from Eq.~\eqref{app:ZFF}.  Figure~\ref{fig:spectrumcheck} compares these spectra with independent diagonalization of Eq.~\eqref{eq:H}.

The comparison covers all nine ordered boundary classes at $N=4,6,8$, with parameters
\begin{equation}
(a_-,a_+,b_-,b_+,\varphi_-,\varphi_+)
=(1.3,0.2,0.7,0.6,0.1,0.6).
\label{app:spectrumparams}
\end{equation}
Each boundary class contains $2^N$ states.  Here $\cos(\varphi_--\varphi_+)>0$, so the $N+1$ modes of the $\ND$--$\ND$ chain obey the $\kappa=+1$ projection; each chain with one nondiagonal end has $N$ independent modes.  The spectra include the boundary strings associated with $a_+=0.2$.  The maximum discrepancy over all $27$ size--boundary combinations is $5.03\times10^{-13}$.

\begin{figure*}[t]
\centering
\includegraphics[width=0.96\textwidth]{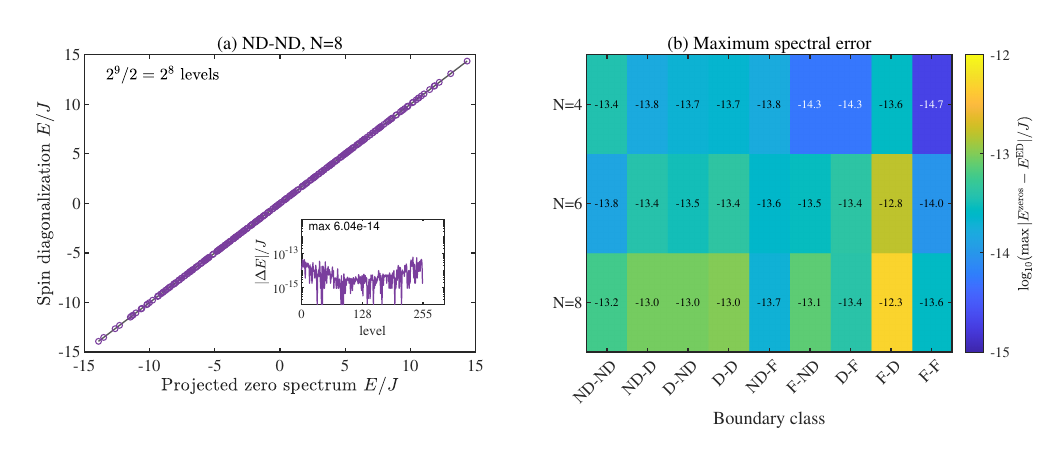}
\caption{Finite-size spectra from the ordinary-zero equation.  (a) All $256$ levels for $\ND$--$\ND$ at $N=8$, compared with exact diagonalization of the spin Hamiltonian.  The inset shows the absolute differences, with a maximum of $6.04\times10^{-14}$.  (b) Maximum absolute spectral differences for all nine ordered boundary classes at $N=4,6,8$.  The maximum over the $27$ cases is $5.03\times10^{-13}$.  Parameters are given in Eq.~\eqref{app:spectrumparams}.}
\label{fig:spectrumcheck}
\end{figure*}

\emph{Diagonal-boundary limit and crossover.---}
The thermodynamic curves follow from the exact series~\eqref{eq:Phiexact} and the real-$b$ integral~\eqref{app:Rkernel}.  Defining $T_{1/2}$ by
\begin{equation}
\frac{\Delta F_{\rm b}(b,T_{1/2})-\mathfrak e(b)}{T_{1/2}}
=\frac14\ln2,
\end{equation}
we obtain $T_{1/2}\ee^{2b}\simeq10.20$ for $b=3$--$6$, consistent with the weak-field scale in Eq.~\eqref{eq:TB}.  Figure~\ref{fig:suppnumerics}(a,b) shows the continuity of the free energy and the final endpoint cancellation.  At fixed $N=10$, Fig.~\ref{fig:suppnumerics}(c) gives $x_{\max}-b/2\to0.824$, whereas the thermodynamic boundary kernel is centered at $x=b+O(1)$.

\begin{figure*}[t]
\centering
\includegraphics[width=0.96\textwidth]{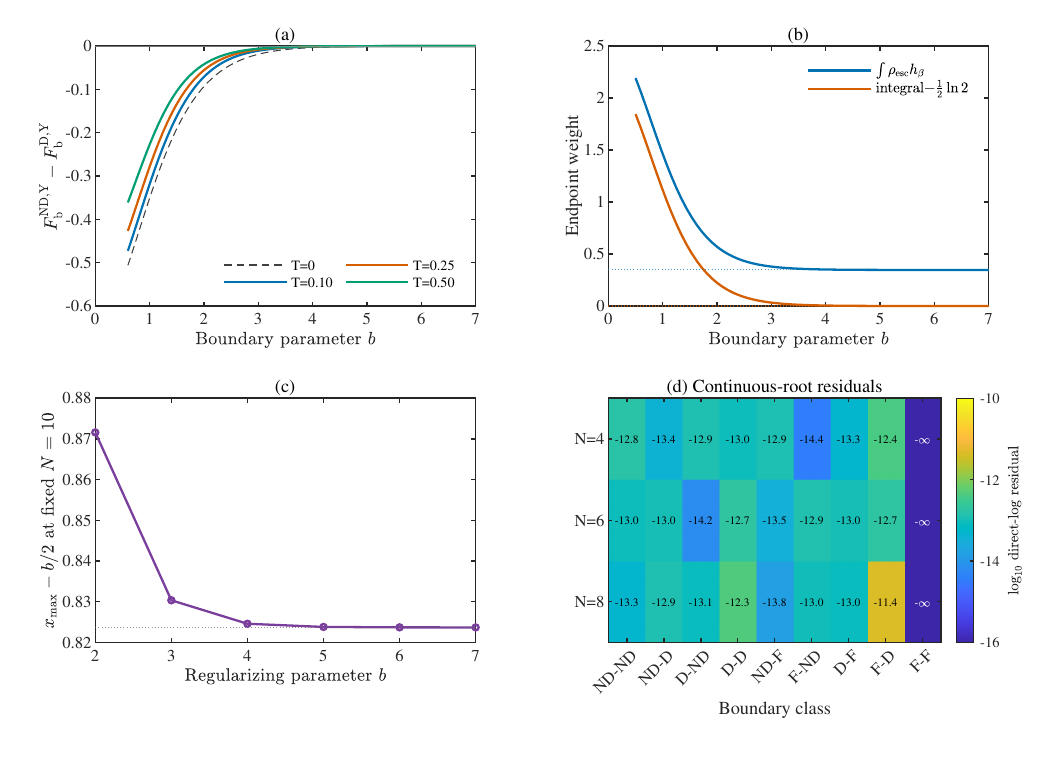}
\caption{Diagonal-boundary limit and finite-size zeros.  (a) The boundary free-energy difference is continuous as $b\to\infty$ at fixed $T>0$, although its linear low-temperature coefficient is singular.  (b) At $\beta=4$, the moving-kernel integral tends to $\frac12\ln2$, canceling the endpoint subtraction.  (c) At fixed $N=10$, $x_{\max}-b/2\to0.824$.  (d) Absolute line-root equation residuals $|\ln|X/R_\nu|+2\pi\ii(Z_\nu-I)|$ for the nine boundary classes and sizes in Fig.~\ref{fig:spectrumcheck}(b).}
\label{fig:suppnumerics}
\end{figure*}

\emph{Finite-size boundary thermodynamics.---}
We compare $\ND$--$\DD$ with $\DD$--$\DD$ at
\begin{equation}
a_-=1.3,\qquad a_+=1.1,\qquad b_-=0.7,\qquad
N=16,32,64,128,256,512,1024,2048,4096 .
\label{app:finiteNparams}
\end{equation}
The boundary angles are $\varphi_-=0.1$ and $\varphi_+=0.6$.  The diagonal reference is the strict $b_-\to+\infty$ limit at fixed $a_-$.  For these parameters, both chains have $N$ independent positive mode energies, no boundary strings, and nondegenerate ground states.  Their ground-state-subtracted partition functions give
\begin{equation}
E_{0,N}^{X}=-\sum_{j=1}^{N}\varepsilon_j^{X},
\qquad
Q_N^{X}(T)=\ln Z_N^{X}+\frac{E_{0,N}^{X}}T
=\sum_{j=1}^{N}\ln(1+\ee^{-2\varepsilon_j^{X}/T}),
\qquad
R_N=Q_N^{\DD,\DD}-Q_N^{\ND,\DD}.
\label{app:stableRN}
\end{equation}
The finite-chain curves in Fig.~\ref{fig:numerics}(c) use all $N$ positive energies obtained from ordinary zeros over $10^{-5}\le T\le0.6$.

As an independent numerical check, the Jordan--Wigner Majoranas can be represented by $\gamma_j=\ii\Gamma_0\Gamma_j$.  The left transverse field and quadratic terms then give
\begin{equation}
H=\frac{i}{4}\boldsymbol{\Gamma}^{\mathsf T}A\boldsymbol{\Gamma},
\qquad
\operatorname{spec}(iA)=\{0,\pm2\varepsilon_1,\ldots,\pm2\varepsilon_N\}.
\label{app:MajoranaCheck}
\end{equation}
Here $A$ is real antisymmetric and has size $2N+1$ for $\ND$--$\DD$; its structural zero adds no physical mode in the $2^N$-dimensional Clifford representation.  The $\DD$--$\DD$ chain uses the usual $2N$-Majorana matrix.  Across all nine sizes, including $N=1024,2048,4096$, and all sampled temperatures, the independent spectra reproduce the ordinary-zero result with $\max|R_N^{\rm zero}-R_N^{\rm Maj}|<3.51\times10^{-9}$.

To quantify proximity to the half-logarithm, define a $5\%$-accuracy interval by
$|R_N-\frac12\ln2|\leq0.05(\frac12\ln2)$.
The intervals and maxima from the zero-derived spectra are
\begin{center}
\begin{tabular}{rccc}
\hline
$N$ & $T_{\max}$ & minimum relative deviation & $5\%$-accuracy interval\\
\hline
1024 & $0.07279$ & $5.886\%$ & none\\
2048 & $0.05141$ & $4.166\%$ & $[0.02756,\,0.09608]$\\
4096 & $0.03633$ & $2.948\%$ & $[0.01184,\,0.11194]$\\
\hline
\end{tabular}
\end{center}
Here $T_{\max}$ locates the maximum of $R_N(T)$, and the minimum relative deviation is $\min_T|R_N(T)-\frac12\ln2|/(\frac12\ln2)$.  All finite-chain curves return to zero at sufficiently low temperature, as expected for nondegenerate ground states.

Figure~\ref{fig:finiteNcheck}(a) shows convergence to the exact finite-temperature $R(b,T)$, with finite-size corrections described by Eq.~\eqref{app:RNcorrection}.  Panel (b) gives the spectral differences between the ordinary-zero and Majorana calculations.  This large-$N$ comparison concerns a single boundary change; the two-end $\ln2$ coefficient is shown in the thermodynamic curves of Fig.~\ref{fig:numerics}(a).

\begin{figure*}[t]
\centering
\includegraphics[width=0.96\textwidth]{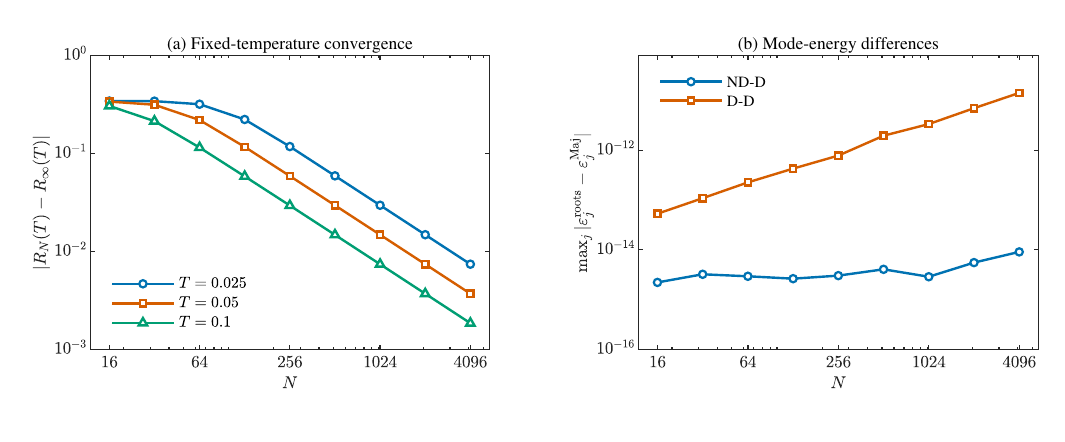}
\caption{Finite-size convergence and spectral accuracy.  Parameters are given in Eq.~\eqref{app:finiteNparams}.  (a) Absolute finite-size difference $|R_N(T)-R(b,T)|$ at three temperatures, with $R(b,T)$ from the exact thermodynamic kernel integral and $R_N$ defined by the same-size ground-state subtraction in Eq.~\eqref{app:RNdef}.  (b) Maximum absolute difference between the sorted positive mode energies obtained from ordinary zeros and independent Majorana matrices.}
\label{fig:finiteNcheck}
\end{figure*}

\end{document}